\documentclass[
 aps,
 prl,
 twocolumn
]{revtex4-2}

\usepackage{lipsum}% http://ctan.org/pkg/lipsum

\usepackage{comment}
\usepackage{graphicx}% Include figure files
\usepackage{dcolumn}% Align table columns on decimal point
\usepackage{bm}% bold math
\usepackage{hyperref}% add hypertext capabilities
\usepackage{color}
\usepackage[caption=false]{subfig}
\usepackage{microtype}
\usepackage{verbatim} 
\usepackage{amsmath,amsfonts,amssymb,amsthm}
\usepackage[algoruled, vlined, lined ,boxed,commentsnumbered]{algorithm2e}

\usepackage{tikz}
\hypersetup{%
        colorlinks=true,
        linkcolor=blue,
        citecolor=blue,
        urlcolor=blue,
      }
\theoremstyle{definition}

\begin{abstract}
As the field of quantum computing progresses to larger-scale devices, multiplexing will be crucial to scale quantum processors. While multiplexed readout is common practice for superconducting devices, relatively little work has been reported about the combination of flux and microwave control lines. Here, we present a method to integrate a microwave line and a flux line into a single
“XYZ line”. 
This combined control line allows us to perform fast single-qubit gates as well as to deliver flux signals to the qubits.
The measured relaxation times of the qubits are comparable to  state-of-art devices employing separate control lines.
We benchmark the fidelity of single-qubit gates with randomized benchmarking, achieving a fidelity above 99.5\%, and we demonstrate that XYZ lines can in principle be used to run parametric entangling gates.
\end{abstract}

\begin{document}

\title{
Full control of superconducting qubits with combined
on-chip microwave and flux lines
}

\author{Riccardo Manenti}
\author{Eyob A. Sete}
\author{Angela Q. Chen}
\author{Shobhan Kulshreshtha}
\author{Jen-Hao Yeh}
\author{Feyza Oruc}
\author{Andrew Bestwick}
\author{Mark Field}
\author{Keith Jackson}
\author{Stefano Poletto}
\affiliation{%
Rigetti Computing,775 Heinz Avenue, Berkeley, CA 94710
}%

\date{\today}

\maketitle

Superconducting quantum processors are one of the leading platforms
for near-term applications and
large-scale quantum computing
due to their flexibility in design, high-fidelity single- and two-qubit gates~\cite{sheldon2016characterizing, rol2017restless, sung2020realization, hong2020demonstration} and fast readout operations~\cite{jeffrey2014fast,
walter2017rapid, heinsoo2018rapid}.
An architecture with a square grid of superconducting qubits for the implementation of the surface code requires each qubit to be coupled to four nearest neighbors as well as individual control lines and resonators~\cite{arute2019quantum, gong2021quantum}. 
Routing the control lines on the perimeter of a monolithic device poses some challenges to scaling  and several techniques have been developed to address this problem, including through-silicon vias and 3D coaxial architectures~\cite{vahidpour2017superconducting, yost2020solid, rahamim2017double}.
3D wiring is not the only challenge for the development of a scalable architecture. As superconducting devices proceed
past the 10-qubit era~\cite{otterbach2017unsupervised, arute2019quantum, gong2021quantum, jurcevic2020demonstration}, 
it becomes necessary to multiplex the waveguides
on the chip. This is already a common practice for readout
lines~\cite{jerger2012frequency,
kelly2015state} but little work has been reported on the combination
of flux (Z) and microwave control (XY) lines
because of the different nature of their coupling to the qubit. The combination of these lines
into a single ``XYZ line'' would approximately halve the number of on-chip input and output ports
for a quantum computer based on tunable qubits, thereby reducing the
complexity of the design and the circuitry in general.

A technical challenge in combining flux and control lines on a planar geometry comes from the  different natures of their coupling to the qubit. The Z line is designed to provide an inductive coupling to the SQUID loop that is high enough to bias the qubit at a specific frequency with a relatively small current such that the overall thermal dissipation does not lead to excessive heating.
This can be done by shorting the Z line to ground approximately $10\,\mu\text{m}$ away from the SQUID loop.
The XY line instead is capacitively coupled to the qubit
pads. 
To minimize the impact on the qubit relaxation time, the capacitive coupling to the $50 \; \Omega$ line cannot be too high~\cite{bardin2021microwaves}. 
At the same time, it cannot be too small otherwise strong microwave signals would be needed to operate single-qubit gates. 
For silicon subtrates with a relative permittivity at low temperature  $\varepsilon_{\text{r}} = 11.45$ \cite{krupka2006measurements}, these requirements are typically satisfied
by positioning the end of the XY line at least $50\;\mu\text{m}$ away from
the nearest qubit pad. 
Hence, combining flux and control lines on the same plane of the qubits can be problematic because of these different interplays. 

In this work, we propose and experimentally demonstrate one possible way to combine  XY and Z lines into a single XYZ line. 
By moving 
the XYZ lines to the surface of the cap using a flip-chip approach~\cite{Vahidpour2017}, the medium separating  the line from the qubit is vacuum whose relative permeability and permittivity are both one. 
This brings the capacitive and inductive coupling between the XYZ line and the qubit on an equal footing. 
We show empirical evidence
that our approach meets the stringent requirements of qubit applications:
keeping a high qubit coherence, supporting microwave drives, and delivering DC and RF flux pulses to the qubits.

\begin{figure}[t]
\includegraphics[width=1\columnwidth]{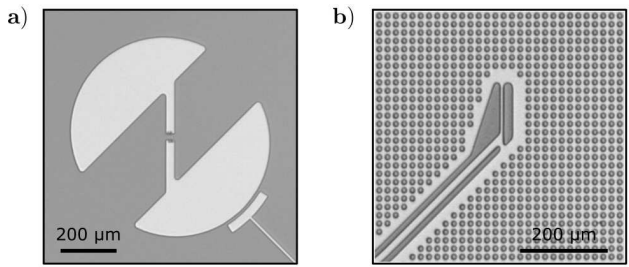}
\caption{
\textbf{A qubit and its associated XYZ line. } 
\textbf{a)} Optical image of one qubit. The silicon substrate is in gray and the niobium film in lighter gray. The two superconducting pads are connected by a SQUID in the center of the image.
\textbf{b)}
Optical image of the corresponding XYZ line on the cap. Almost the entire surface is covered by a meshed ground plane.
}
\label{fig:quic-cap}
\end{figure}

The device used for this investigation includes four tunable qubits, each capacitively coupled to a readout resonator.
The cap contains two separate readout lines (each capacitively coupled to two resonators) and four XYZ lines, one per qubit. Most of the cap surface is covered with a  meshed ground plane. 
The region of the cap that surmounts the qubits and resonators is characterized by a $24\;\mu\text{m}$ deep cavity~\cite{o2017superconducting}. The device is bonded to a $6\times 6 \text{ mm}^2$ cap with a flip-chip bonder that provides an alignment precision of a few micrometers. 
Since the
height of the flattened indium bumps
is about $3\;\mu\text{m}$, the distance between the qubit and the ground plane above is  $27\;\mu\text{m}$ (see Fig.~\ref{fig:quic-cap} for an
an  optical image of the qubit and the associated XYZ line on top).
The electrical connection between the two chips is tested at room temperature with dedicated test structures.
The cap is wirebonded to a printed circuit board  and mounted to the coldest plate of a dilution refrigerator with a $10\;\text{mK}$ base temperature.

The attenuation and filtering of the  XY and Z fridge lines connecting the room temperature instrumentation to the device are different. This is because the XY lines  must support microwave signals in the $\text{3-7 GHz}$ band, which is the typical qubit frequency band. In addition, these lines must have a strong attenuation to reduce the thermal noise reaching the device~\cite{krinner2019engineering}. The Z lines instead
provide low frequency pulses ($\text{DC}-1.5 \text{ GHz}$) in order to bias the qubits at specific sweet spots and operate parametric entangling gates~\cite{caldwell2018parametrically}. The Z lines require a smaller attenuation than the XY lines.

\begin{figure}[t]
\includegraphics[width=1\columnwidth]{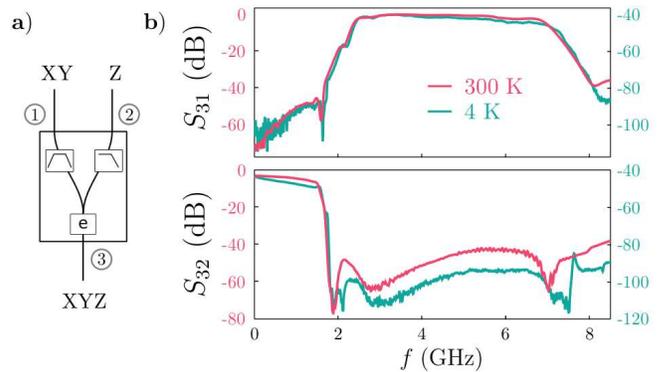}
\caption{
\textbf{The diplexer. } 
\textbf{a)} Diagrammatic representation of the cryogenic diplexer. The input port 1 (2) receives 3-7 GHz (DC-1 GHz) microwave signals. They get combined and routed to port 3. The diplexer includes a low pass filter, a band pass filter and an eccosorb.
\textbf{b)}
Measured transmission coefficient between ports $1\rightarrow 3$ and $2\rightarrow 3$ for a typical diplexer at room temperature (magenta curve) and at 4~K (teal curve). The cryogenic measurements were taken with a 40 dB attenuation between the VNA and the diplexer inside the cryostat. 
}
\label{fig:diplexer}
\end{figure}

 Due to the different filtering requirements, the XY and Z fridge lines are combined at the lowest temperature plate of the dilution refrigerator. To this end, we have developed an in-house cryogenic  diplexer as shown in Fig.~\ref{fig:diplexer}a.  This device  allows us not only to combine low and high frequency signals but also to filter the frequency components outside a specific frequency band by using a network of inductors and capacitors. 
To mitigate the injection of quasi particles generated by high-frequency photons, the diplexer also includes an eccosorb filter on the output line~\cite{corcoles2011protecting, barends2011minimizing}. 
Figure~\ref{fig:diplexer}b shows the measured transmission coefficient of a typical diplexer at room temperature and at $4\;\text{K}$. 
The diplexer was designed to have 
a \text{3-7 GHz} bandpass filter for the XY line (port 1)
and a \text{1.5 GHz}  
low-pass filter for the Z line (port 2).
The transmission coefficient from port 1 to port 2  (not shown in the figure) is lower than $-\text{20 dB}$ up to $15 \text{ GHz}$.  In our experiment, 
one diplexer per qubit is thermally anchored to the 10 mK stage of the dilution refrigerator.
We have performed several cool downs and we have not noticed any degradation of their functionalities.

 The geometry of the XYZ line has been engineered to 
 obtain an upper limit on $T_1$ greater than $\text{200 } \mu \text{s}$, a mutual inductance with the SQUID loop of approximately $500 \text{ nH}$ and a capacitive coupling high enough to enable $20 \text{ ns}$ $\pi$-pulses with our fridge setup (see Supplementary Information for a schematic of the fridge).  
 To optimize these design parameters, we have performed 
 full wave microwave simulations.
 %simulations with  Ansys HFSS.
 The mutual inductance was tuned by varying the distance between the XYZ line and the SQUID as well as the width and length of the inductors shorting the line to ground. The mutual inductance can also be adjusted by changing  the area of the SQUID loop. However, 
 its perimeter cannot be too long in order not to limit the qubit coherence~\cite{braumuller2020characterizing}.
 With regards to the capacitive coupling, 
 the area of the XYZ line that surmounts the qubit pads affects the coupling strength. As a result, the capacitive coupling can vary depending on the qubit geometry.
 The geometry illustrated in Fig.~\ref{fig:quic-cap}b is the result of an optimization process that takes into account the qubit relaxation time and the capacitive and inductive couplings.

 Characterization of the device at base temperature shows that we can tune the qubit frequencies between 
$3.0 - 3.8\;\text{GHz}$.
The median of the relaxation time for all of the qubits over a day  is $\tilde T_1 = 53\;\mu \text{s}$, and 
% hereas the median of the 
transverse relaxation times at the maximum qubit frequency is
$\tilde T^*_{2} = 10\;\mu \text{s}$ and $\tilde T_{\text{2E}} = 49\;\mu \text{s}$. 
The relaxation times fluctuate over time (see   Supplementary Information).
This phenomenon has been reported elsewhere~\cite{burnett2019decoherence, schlor2019correlating}.

\begin{figure}[t]
\includegraphics[width=1\columnwidth]{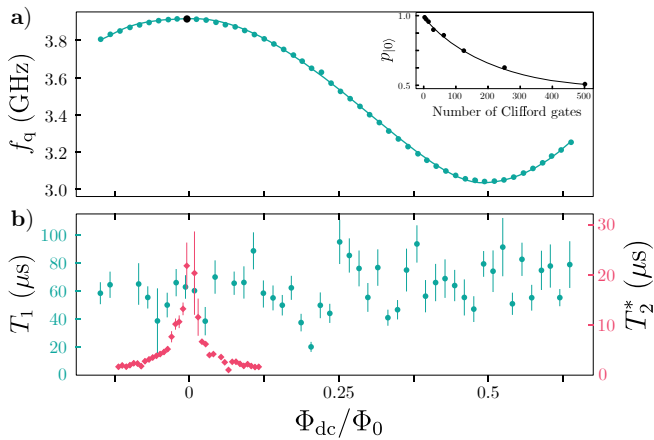}
\caption{
\textbf{
Characterization.}
\textbf{a)} Qubit frequency as a function of flux bias. The black dot represents $f_{\text{max}}$. 
Inset: 
randomized benchmarking with optimized 100 ns single-qubit gates run at $f_{\text{max}}$. The solid line is a fit to $Ap^n + B$. From the fit, we extract a single-qubit gate fidelity of $99.77 \pm 0.02 \%$.
\textbf{b)} Measured relaxation time $T_1$ (teal dots) and transverse relaxation time $T^*_2$ (magenta diamonds) as a function of flux bias. The error bars represent the uncertainty of the exponential fit for each experiment. 
}
\label{fig:rabi-rb}
\end{figure}

 We first verify the functionality of the XY lines by testing their ability to manipulate the qubit state.
 After some initial Rabi experiments, we optimize \text{100 ns} DRAG Gaussian pulses and measure a single-qubit gate fidelity of  $99.77 \pm 0.02 \%$
  with randomized benchmarking for qubit~1 (see the inset of Fig.~\ref{fig:rabi-rb}a). The theoretical limit imposed by the relaxation time of this qubit is $99.84\%$~\cite{asaad2016independent}. This demonstrates that 
 high-fidelity single-qubit gates can be implemented with the XYZ lines. 
 
 We first assess 
  that the XYZ lines can be used to deliver flux signals to the qubits. 
 We measure the qubit frequency $f_{\text{q}}$ as a function of the applied DC current as shown in Fig.~\ref{fig:rabi-rb}a.
 The qubit frequency is measured with spectroscopic measurements and Ramsey experiments. The data points are fitted with an analytical transmon model. 
 At each flux bias, we measure the relaxation time $T_1$ (see Fig.~\ref{fig:rabi-rb}b).
 The value of $T_1$ does not show a significant flux dependence and its average value is $75\; \mu \text{s}$. 
 Figure~\ref{fig:rabi-rb}b includes the measurement of $T^*_{\text{2}}$ as a function of flux close to the DC sweet spot. 
 As expected, the transverse relaxation time increases substantially at the sweet spot where the sensitivity of the qubit frequency to flux noise is the lowest.
 Close to the DC sweet spots,  $T^*_{\text{2}}$ is above $10\; \mu\text{s}$ allowing high-fidelity single-qubit gates. Other devices fabricated on the same wafer show similar performance although they were measured without XYZ lines.
 We can thus conclude that XYZ lines can be used to tune the qubit frequency without compromising their relaxation time. 
 
 \begin{figure}[t]
\includegraphics[width=1\columnwidth]{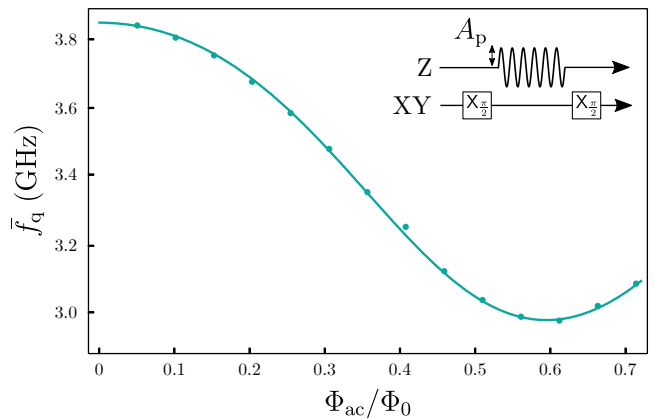}
\caption{
\textbf{Ramsey experiment with flux pulses.} 
Time average qubit frequency as a function of flux pulse amplitude in a Ramsey type experiment. The qubit is initially parked at the maximum frequency $f_{\text{max}}$. After the first $\pi/2$ pulse, a flux pulse with amplitude $A_{\text{p}}$ is applied to the qubit. This pulse makes the qubit frequency oscillate about $f_{\text{max}}$. This changes the effective qubit frequency to $\overline{f}_{\text q}$. The solid curve is a fit to Eq.~\eqref{eq:modulation-detuning-eq}. The amplitude has been expressed in terms of flux quanta using $\Phi_{\text{ac}} = \beta A_{\text p}$ where the factor $\beta = 0.510 \;\Phi_0/\text V$ has been extracted from a fit.
}
\label{fig:stark-shift}
\end{figure}
 
 Next, we validate the ability of the XYZ lines to support RF flux pulses, crucial for the implementation of parametric entangling gates~\cite{caldwell2018parametrically, mckay2016universal}.
 In our experiment, the qubit starts at its maximum frequency $f_{\text{max}}$. A flux RF pulse 
 $\Phi(t) = \Phi_{\text{dc}} + \Phi_{\text{ac}} \cos(\omega_\text{d}t)$  is delivered to the qubit where $\Phi_{\text{dc}}$ is the DC bias, $\Phi_{\text{ac}}$ is the flux pulse amplitude and $\omega_\text{d}$ is the pulse frequency. 
The flux pulse
induces periodic oscillations of the qubit frequency.
 The time average qubit frequency is measured with a Ramsey type experiment with the parametric modulation applied between two $\pi/2$ pulses.
 Figure~\ref{fig:stark-shift} shows the measured effective qubit frequency as a function of the flux pulse amplitude. The data points are fitted to~\cite{Didier2017}
 \begin{eqnarray}
 \bar{f}_{\text{q}}(\Phi_{\text{ac}})&=& 
 \frac{1}{T}\int^T_0 f_{\text{q}}(t')dt'
 \nonumber \\
 &=&\sum_{n\geq 0}^{\infty}s_{n}\cos(2\pi n\Phi_{\text{dc}})J_{0}(2\pi n \Phi_{\text{ac}})
 \label{eq:modulation-detuning-eq}
 \end{eqnarray}
 where $T$ is the oscillation period, $J_0(x)$ is the Bessel function of the first kind, 
 and the constants $s_n$ only depend on the Josephson and charging energies $E_{\text{J1}}$, $E_{\text{J2}}$, and $E_{\text{C}}$ (their analytical expression is presented in the Supplementary Information). 
 In our experiment, $\Phi_{\text{dc}}=0$ since the qubit is initially parked at $f_{\text{max}}$.
 The flux pulse amplitude can be expressed in terms of the amplitude $A_{\text{p}}$ generated by the room-temperature instrumentation as $\Phi_{\text{ac}} = \beta A_{\text{p}}$ where $\beta$ is a factor that can be extracted  from the fit.
 As shown in the figure,  
 we were able to reach the AC sweet spot, $f_{\text{min}}$, where we operate parametric entangling gates \cite{Didier2019}. 
 Devices with XYZ lines are now used in our lab to run parametric entangling gates routinely.

The combination of the XY and Z fridge lines into a single line by means of a cryogenic diplexer 
may lead to an undesired effect.
When a $\pi$ pulse is sent through the XY fridge line to excite the qubit, it produces a current that flows through the termination of the XYZ line. This current can inadvertently  modulate the qubit frequency.
In our setup, a 100 ns $\pi$ pulse is implemented with a room-temperature amplitude of $V_{\text p}=0.3\; \text{V}$. The signal reaching the device creates a magnetic field through the SQUID of $\Phi_{\text{ac}} = 1.6\cdot 10^{-4}\Phi_0$ (here,  we assumed that the attenuation of the line is $85\;\text{dB}$ at the qubit frequency and the mutual inductance between the XYZ line and the SQUID is $M = 500\;\text{fH}$). This flux  does not modulate the qubit frequency by an appreciable amount. Indeed, by approximating Eq. (1) close to $f_{\text{max}}$ up to second order in $\Phi_{\text{ac}}$, we obtain
\begin{equation}
\delta f = \bar{f}_{\text q}-f_{\text {max}} = 
-
\frac{\pi^2 E_{\text J 1}/E_{\text J2}}{2h(1+E_{\text J1}/E_{\text J2})^2} \sqrt{8 E_{\text J\Sigma} 
E_{\text{C}}}
\left( \frac{\Phi_{\text{ac}}}{\Phi_0}\right)^2
\end{equation}
and using the parameters for qubit 0 
($E_{\text{J1}}/h = 2140\;\text{MHz}$, $E_{\text{J2}}/h = 9040\;\text{MHz}$, 
$E_{\text{C}}/h = 182\;\text{MHz}$, 
$E_{\text J\Sigma} =E_{\text{J1}}+E_{\text{J2}} $), we obtain a frequency shift of $\delta f = -\text{79 Hz}$.
This shift is  below the qubit linewidth and cannot be detected with Ramsey experiments.

In conclusion, we demonstrated full control of superconducting qubits by combining XY and Z lines into a single XYZ line.
We showed that XYZ lines
patterned on the surface of the cap
can be used to implement  fast single qubit-gates and tune the qubit frequency with both DC and fast flux biases. We tested the performance of single-qubit gates with  randomized benchmarking 
achieving a fidelity as high as $99.93\pm 0.04\%$.
Realization of combined XY and Z lines can reduce the number of on-chip input ports, an important requirement when scaling superconducting quantum processors. The natural next step is to combine the XY cables and the Z cables into a single cable to further reduce the complexity of the fridge built out for quantum processors with a large number of qubits. This can be accomplished by engineering frequency-dependent attenuators. More sophisticated  techniques for the  multiplexing of the cables  would require optical links \cite{lecocq2021control}.

\section*{Acknowledgments}

This material is based upon work supported by Rigetti Computing and the Defense Advanced Research Projects Agency (DARPA) under agreement No. HR00112090058.

\section*{Contributions}

R.M. and S.P. developed the proposal. R.M., E.A.S., and S.P. acquired the data. R.M., E.A.S. performed the data analysis. R.M. simulated and designed the device and the cap. J.-H.Y. simulated and designed the diplexer. F.O., A.B., M.F. and K.J. were responsible for the development of an optimal cap fabrication. R.M. wrote the manuscript. E.A.S., A.C., S. K., and S.P. edited the manuscript. S.P. was the principal investigator of the effort.

\bibliography{XYZ_bib}

%apsrev4-2.bst 2019-01-14 (MD) hand-edited version of apsrev4-1.bst
%Control: key (0)
%Control: author (8) initials jnrlst
%Control: editor formatted (1) identically to author
%Control: production of article title (0) allowed
%Control: page (0) single
%Control: year (1) truncated
%Control: production of eprint (0) enabled
\begin{thebibliography}{32}%
\makeatletter
\providecommand \@ifxundefined [1]{%
 \@ifx{#1\undefined}
}%
\providecommand \@ifnum [1]{%
 \ifnum #1\expandafter \@firstoftwo
 \else \expandafter \@secondoftwo
 \fi
}%
\providecommand \@ifx [1]{%
 \ifx #1\expandafter \@firstoftwo
 \else \expandafter \@secondoftwo
 \fi
}%
\providecommand \natexlab [1]{#1}%
\providecommand \enquote  [1]{``#1''}%
\providecommand \bibnamefont  [1]{#1}%
\providecommand \bibfnamefont [1]{#1}%
\providecommand \citenamefont [1]{#1}%
\providecommand \href@noop [0]{\@secondoftwo}%
\providecommand \href [0]{\begingroup \@sanitize@url \@href}%
\providecommand \@href[1]{\@@startlink{#1}\@@href}%
\providecommand \@@href[1]{\endgroup#1\@@endlink}%
\providecommand \@sanitize@url [0]{\catcode `\\12\catcode `\$12\catcode
  `\&12\catcode `\#12\catcode `\^12\catcode `\_12\catcode `\%12\relax}%
\providecommand \@@startlink[1]{}%
\providecommand \@@endlink[0]{}%
\providecommand \url  [0]{\begingroup\@sanitize@url \@url }%
\providecommand \@url [1]{\endgroup\@href {#1}{\urlprefix }}%
\providecommand \urlprefix  [0]{URL }%
\providecommand \Eprint [0]{\href }%
\providecommand \doibase [0]{https://doi.org/}%
\providecommand \selectlanguage [0]{\@gobble}%
\providecommand \bibinfo  [0]{\@secondoftwo}%
\providecommand \bibfield  [0]{\@secondoftwo}%
\providecommand \translation [1]{[#1]}%
\providecommand \BibitemOpen [0]{}%
\providecommand \bibitemStop [0]{}%
\providecommand \bibitemNoStop [0]{.\EOS\space}%
\providecommand \EOS [0]{\spacefactor3000\relax}%
\providecommand \BibitemShut  [1]{\csname bibitem#1\endcsname}%
\let\auto@bib@innerbib\@empty
%</preamble>
\bibitem [{\citenamefont {Sheldon}\ \emph {et~al.}(2016)\citenamefont
  {Sheldon}, \citenamefont {Bishop}, \citenamefont {Magesan}, \citenamefont
  {Filipp}, \citenamefont {Chow},\ and\ \citenamefont
  {Gambetta}}]{sheldon2016characterizing}%
  \BibitemOpen
  \bibfield  {author} {\bibinfo {author} {\bibfnamefont {S.}~\bibnamefont
  {Sheldon}}, \bibinfo {author} {\bibfnamefont {L.~S.}\ \bibnamefont {Bishop}},
  \bibinfo {author} {\bibfnamefont {E.}~\bibnamefont {Magesan}}, \bibinfo
  {author} {\bibfnamefont {S.}~\bibnamefont {Filipp}}, \bibinfo {author}
  {\bibfnamefont {J.~M.}\ \bibnamefont {Chow}},\ and\ \bibinfo {author}
  {\bibfnamefont {J.~M.}\ \bibnamefont {Gambetta}},\ }\bibfield  {title}
  {\bibinfo {title} {Characterizing errors on qubit operations via iterative
  randomized benchmarking},\ }\href@noop {} {\bibfield  {journal} {\bibinfo
  {journal} {Physical Review A}\ }\textbf {\bibinfo {volume} {93}},\ \bibinfo
  {pages} {012301} (\bibinfo {year} {2016})}\BibitemShut {NoStop}%
\bibitem [{\citenamefont {Rol}\ \emph {et~al.}(2017)\citenamefont {Rol},
  \citenamefont {Bultink}, \citenamefont {O’Brien}, \citenamefont {De~Jong},
  \citenamefont {Theis}, \citenamefont {Fu}, \citenamefont {Luthi},
  \citenamefont {Vermeulen}, \citenamefont {de~Sterke}, \citenamefont {Bruno}
  \emph {et~al.}}]{rol2017restless}%
  \BibitemOpen
  \bibfield  {author} {\bibinfo {author} {\bibfnamefont {M.}~\bibnamefont
  {Rol}}, \bibinfo {author} {\bibfnamefont {C.}~\bibnamefont {Bultink}},
  \bibinfo {author} {\bibfnamefont {T.}~\bibnamefont {O’Brien}}, \bibinfo
  {author} {\bibfnamefont {S.}~\bibnamefont {De~Jong}}, \bibinfo {author}
  {\bibfnamefont {L.}~\bibnamefont {Theis}}, \bibinfo {author} {\bibfnamefont
  {X.}~\bibnamefont {Fu}}, \bibinfo {author} {\bibfnamefont {F.}~\bibnamefont
  {Luthi}}, \bibinfo {author} {\bibfnamefont {R.}~\bibnamefont {Vermeulen}},
  \bibinfo {author} {\bibfnamefont {J.}~\bibnamefont {de~Sterke}}, \bibinfo
  {author} {\bibfnamefont {A.}~\bibnamefont {Bruno}}, \emph {et~al.},\
  }\bibfield  {title} {\bibinfo {title} {Restless tuneup of high-fidelity qubit
  gates},\ }\href@noop {} {\bibfield  {journal} {\bibinfo  {journal} {Physical
  Review Applied}\ }\textbf {\bibinfo {volume} {7}},\ \bibinfo {pages} {041001}
  (\bibinfo {year} {2017})}\BibitemShut {NoStop}%
\bibitem [{\citenamefont {Sung}\ \emph {et~al.}(2020)\citenamefont {Sung},
  \citenamefont {Ding}, \citenamefont {Braum{\"u}ller}, \citenamefont
  {Veps{\"a}l{\"a}inen}, \citenamefont {Kannan}, \citenamefont {Kjaergaard},
  \citenamefont {Greene}, \citenamefont {Samach}, \citenamefont {McNally},
  \citenamefont {Kim} \emph {et~al.}}]{sung2020realization}%
  \BibitemOpen
  \bibfield  {author} {\bibinfo {author} {\bibfnamefont {Y.}~\bibnamefont
  {Sung}}, \bibinfo {author} {\bibfnamefont {L.}~\bibnamefont {Ding}}, \bibinfo
  {author} {\bibfnamefont {J.}~\bibnamefont {Braum{\"u}ller}}, \bibinfo
  {author} {\bibfnamefont {A.}~\bibnamefont {Veps{\"a}l{\"a}inen}}, \bibinfo
  {author} {\bibfnamefont {B.}~\bibnamefont {Kannan}}, \bibinfo {author}
  {\bibfnamefont {M.}~\bibnamefont {Kjaergaard}}, \bibinfo {author}
  {\bibfnamefont {A.}~\bibnamefont {Greene}}, \bibinfo {author} {\bibfnamefont
  {G.~O.}\ \bibnamefont {Samach}}, \bibinfo {author} {\bibfnamefont
  {C.}~\bibnamefont {McNally}}, \bibinfo {author} {\bibfnamefont
  {D.}~\bibnamefont {Kim}}, \emph {et~al.},\ }\bibfield  {title} {\bibinfo
  {title} {Realization of high-fidelity cz and zz-free iswap gates with a
  tunable coupler},\ }\href@noop {} {\bibfield  {journal} {\bibinfo  {journal}
  {arXiv preprint arXiv:2011.01261}\ } (\bibinfo {year} {2020})}\BibitemShut
  {NoStop}%
\bibitem [{\citenamefont {Hong}\ \emph {et~al.}(2020)\citenamefont {Hong},
  \citenamefont {Papageorge}, \citenamefont {Sivarajah}, \citenamefont
  {Crossman}, \citenamefont {Didier}, \citenamefont {Polloreno}, \citenamefont
  {Sete}, \citenamefont {Turkowski}, \citenamefont {da~Silva},\ and\
  \citenamefont {Johnson}}]{hong2020demonstration}%
  \BibitemOpen
  \bibfield  {author} {\bibinfo {author} {\bibfnamefont {S.~S.}\ \bibnamefont
  {Hong}}, \bibinfo {author} {\bibfnamefont {A.~T.}\ \bibnamefont
  {Papageorge}}, \bibinfo {author} {\bibfnamefont {P.}~\bibnamefont
  {Sivarajah}}, \bibinfo {author} {\bibfnamefont {G.}~\bibnamefont {Crossman}},
  \bibinfo {author} {\bibfnamefont {N.}~\bibnamefont {Didier}}, \bibinfo
  {author} {\bibfnamefont {A.~M.}\ \bibnamefont {Polloreno}}, \bibinfo {author}
  {\bibfnamefont {E.~A.}\ \bibnamefont {Sete}}, \bibinfo {author}
  {\bibfnamefont {S.~W.}\ \bibnamefont {Turkowski}}, \bibinfo {author}
  {\bibfnamefont {M.~P.}\ \bibnamefont {da~Silva}},\ and\ \bibinfo {author}
  {\bibfnamefont {B.~R.}\ \bibnamefont {Johnson}},\ }\bibfield  {title}
  {\bibinfo {title} {Demonstration of a parametrically activated entangling
  gate protected from flux noise},\ }\href@noop {} {\bibfield  {journal}
  {\bibinfo  {journal} {Physical Review A}\ }\textbf {\bibinfo {volume}
  {101}},\ \bibinfo {pages} {012302} (\bibinfo {year} {2020})}\BibitemShut
  {NoStop}%
\bibitem [{\citenamefont {Jeffrey}\ \emph {et~al.}(2014)\citenamefont
  {Jeffrey}, \citenamefont {Sank}, \citenamefont {Mutus}, \citenamefont
  {White}, \citenamefont {Kelly}, \citenamefont {Barends}, \citenamefont
  {Chen}, \citenamefont {Chen}, \citenamefont {Chiaro}, \citenamefont
  {Dunsworth} \emph {et~al.}}]{jeffrey2014fast}%
  \BibitemOpen
  \bibfield  {author} {\bibinfo {author} {\bibfnamefont {E.}~\bibnamefont
  {Jeffrey}}, \bibinfo {author} {\bibfnamefont {D.}~\bibnamefont {Sank}},
  \bibinfo {author} {\bibfnamefont {J.}~\bibnamefont {Mutus}}, \bibinfo
  {author} {\bibfnamefont {T.}~\bibnamefont {White}}, \bibinfo {author}
  {\bibfnamefont {J.}~\bibnamefont {Kelly}}, \bibinfo {author} {\bibfnamefont
  {R.}~\bibnamefont {Barends}}, \bibinfo {author} {\bibfnamefont
  {Y.}~\bibnamefont {Chen}}, \bibinfo {author} {\bibfnamefont {Z.}~\bibnamefont
  {Chen}}, \bibinfo {author} {\bibfnamefont {B.}~\bibnamefont {Chiaro}},
  \bibinfo {author} {\bibfnamefont {A.}~\bibnamefont {Dunsworth}}, \emph
  {et~al.},\ }\bibfield  {title} {\bibinfo {title} {Fast accurate state
  measurement with superconducting qubits},\ }\href@noop {} {\bibfield
  {journal} {\bibinfo  {journal} {Phys. Rev. Lett.}\ }\textbf {\bibinfo
  {volume} {112}},\ \bibinfo {pages} {190504} (\bibinfo {year}
  {2014})}\BibitemShut {NoStop}%
\bibitem [{\citenamefont {Walter}\ \emph {et~al.}(2017)\citenamefont {Walter},
  \citenamefont {Kurpiers}, \citenamefont {Gasparinetti}, \citenamefont
  {Magnard}, \citenamefont {Poto{\v{c}}nik}, \citenamefont {Salath{\'e}},
  \citenamefont {Pechal}, \citenamefont {Mondal}, \citenamefont {Oppliger},
  \citenamefont {Eichler} \emph {et~al.}}]{walter2017rapid}%
  \BibitemOpen
  \bibfield  {author} {\bibinfo {author} {\bibfnamefont {T.}~\bibnamefont
  {Walter}}, \bibinfo {author} {\bibfnamefont {P.}~\bibnamefont {Kurpiers}},
  \bibinfo {author} {\bibfnamefont {S.}~\bibnamefont {Gasparinetti}}, \bibinfo
  {author} {\bibfnamefont {P.}~\bibnamefont {Magnard}}, \bibinfo {author}
  {\bibfnamefont {A.}~\bibnamefont {Poto{\v{c}}nik}}, \bibinfo {author}
  {\bibfnamefont {Y.}~\bibnamefont {Salath{\'e}}}, \bibinfo {author}
  {\bibfnamefont {M.}~\bibnamefont {Pechal}}, \bibinfo {author} {\bibfnamefont
  {M.}~\bibnamefont {Mondal}}, \bibinfo {author} {\bibfnamefont
  {M.}~\bibnamefont {Oppliger}}, \bibinfo {author} {\bibfnamefont
  {C.}~\bibnamefont {Eichler}}, \emph {et~al.},\ }\bibfield  {title} {\bibinfo
  {title} {Rapid high-fidelity single-shot dispersive readout of
  superconducting qubits},\ }\href@noop {} {\bibfield  {journal} {\bibinfo
  {journal} {Physical Review Applied}\ }\textbf {\bibinfo {volume} {7}},\
  \bibinfo {pages} {054020} (\bibinfo {year} {2017})}\BibitemShut {NoStop}%
\bibitem [{\citenamefont {Heinsoo}\ \emph {et~al.}(2018)\citenamefont
  {Heinsoo}, \citenamefont {Andersen}, \citenamefont {Remm}, \citenamefont
  {Krinner}, \citenamefont {Walter}, \citenamefont {Salath{\'e}}, \citenamefont
  {Gasparinetti}, \citenamefont {Besse}, \citenamefont {Poto{\v{c}}nik},
  \citenamefont {Wallraff} \emph {et~al.}}]{heinsoo2018rapid}%
  \BibitemOpen
  \bibfield  {author} {\bibinfo {author} {\bibfnamefont {J.}~\bibnamefont
  {Heinsoo}}, \bibinfo {author} {\bibfnamefont {C.~K.}\ \bibnamefont
  {Andersen}}, \bibinfo {author} {\bibfnamefont {A.}~\bibnamefont {Remm}},
  \bibinfo {author} {\bibfnamefont {S.}~\bibnamefont {Krinner}}, \bibinfo
  {author} {\bibfnamefont {T.}~\bibnamefont {Walter}}, \bibinfo {author}
  {\bibfnamefont {Y.}~\bibnamefont {Salath{\'e}}}, \bibinfo {author}
  {\bibfnamefont {S.}~\bibnamefont {Gasparinetti}}, \bibinfo {author}
  {\bibfnamefont {J.-C.}\ \bibnamefont {Besse}}, \bibinfo {author}
  {\bibfnamefont {A.}~\bibnamefont {Poto{\v{c}}nik}}, \bibinfo {author}
  {\bibfnamefont {A.}~\bibnamefont {Wallraff}}, \emph {et~al.},\ }\bibfield
  {title} {\bibinfo {title} {Rapid high-fidelity multiplexed readout of
  superconducting qubits},\ }\href@noop {} {\bibfield  {journal} {\bibinfo
  {journal} {Physical Review Applied}\ }\textbf {\bibinfo {volume} {10}},\
  \bibinfo {pages} {034040} (\bibinfo {year} {2018})}\BibitemShut {NoStop}%
\bibitem [{\citenamefont {Arute}\ \emph {et~al.}(2019)\citenamefont {Arute},
  \citenamefont {Arya}, \citenamefont {Babbush}, \citenamefont {Bacon},
  \citenamefont {Bardin}, \citenamefont {Barends}, \citenamefont {Biswas},
  \citenamefont {Boixo}, \citenamefont {Brandao}, \citenamefont {Buell} \emph
  {et~al.}}]{arute2019quantum}%
  \BibitemOpen
  \bibfield  {author} {\bibinfo {author} {\bibfnamefont {F.}~\bibnamefont
  {Arute}}, \bibinfo {author} {\bibfnamefont {K.}~\bibnamefont {Arya}},
  \bibinfo {author} {\bibfnamefont {R.}~\bibnamefont {Babbush}}, \bibinfo
  {author} {\bibfnamefont {D.}~\bibnamefont {Bacon}}, \bibinfo {author}
  {\bibfnamefont {J.~C.}\ \bibnamefont {Bardin}}, \bibinfo {author}
  {\bibfnamefont {R.}~\bibnamefont {Barends}}, \bibinfo {author} {\bibfnamefont
  {R.}~\bibnamefont {Biswas}}, \bibinfo {author} {\bibfnamefont
  {S.}~\bibnamefont {Boixo}}, \bibinfo {author} {\bibfnamefont {F.~G.}\
  \bibnamefont {Brandao}}, \bibinfo {author} {\bibfnamefont {D.~A.}\
  \bibnamefont {Buell}}, \emph {et~al.},\ }\bibfield  {title} {\bibinfo {title}
  {Quantum supremacy using a programmable superconducting processor},\
  }\href@noop {} {\bibfield  {journal} {\bibinfo  {journal} {Nature}\ }\textbf
  {\bibinfo {volume} {574}},\ \bibinfo {pages} {505} (\bibinfo {year}
  {2019})}\BibitemShut {NoStop}%
\bibitem [{\citenamefont {Gong}\ \emph {et~al.}(2021)\citenamefont {Gong},
  \citenamefont {Wang}, \citenamefont {Zha}, \citenamefont {Chen},
  \citenamefont {Huang}, \citenamefont {Wu}, \citenamefont {Zhu}, \citenamefont
  {Zhao}, \citenamefont {Li}, \citenamefont {Guo} \emph
  {et~al.}}]{gong2021quantum}%
  \BibitemOpen
  \bibfield  {author} {\bibinfo {author} {\bibfnamefont {M.}~\bibnamefont
  {Gong}}, \bibinfo {author} {\bibfnamefont {S.}~\bibnamefont {Wang}}, \bibinfo
  {author} {\bibfnamefont {C.}~\bibnamefont {Zha}}, \bibinfo {author}
  {\bibfnamefont {M.-C.}\ \bibnamefont {Chen}}, \bibinfo {author}
  {\bibfnamefont {H.-L.}\ \bibnamefont {Huang}}, \bibinfo {author}
  {\bibfnamefont {Y.}~\bibnamefont {Wu}}, \bibinfo {author} {\bibfnamefont
  {Q.}~\bibnamefont {Zhu}}, \bibinfo {author} {\bibfnamefont {Y.}~\bibnamefont
  {Zhao}}, \bibinfo {author} {\bibfnamefont {S.}~\bibnamefont {Li}}, \bibinfo
  {author} {\bibfnamefont {S.}~\bibnamefont {Guo}}, \emph {et~al.},\ }\bibfield
   {title} {\bibinfo {title} {Quantum walks on a programmable two-dimensional
  62-qubit superconducting processor},\ }\href@noop {} {\bibfield  {journal}
  {\bibinfo  {journal} {arXiv preprint arXiv:2102.02573}\ } (\bibinfo {year}
  {2021})}\BibitemShut {NoStop}%
\bibitem [{\citenamefont {Vahidpour}\ \emph
  {et~al.}(2017{\natexlab{a}})\citenamefont {Vahidpour}, \citenamefont
  {O'Brien}, \citenamefont {Whyland}, \citenamefont {Angeles}, \citenamefont
  {Marshall}, \citenamefont {Scarabelli}, \citenamefont {Crossman},
  \citenamefont {Yadav}, \citenamefont {Mohan}, \citenamefont {Bui} \emph
  {et~al.}}]{vahidpour2017superconducting}%
  \BibitemOpen
  \bibfield  {author} {\bibinfo {author} {\bibfnamefont {M.}~\bibnamefont
  {Vahidpour}}, \bibinfo {author} {\bibfnamefont {W.}~\bibnamefont {O'Brien}},
  \bibinfo {author} {\bibfnamefont {J.~T.}\ \bibnamefont {Whyland}}, \bibinfo
  {author} {\bibfnamefont {J.}~\bibnamefont {Angeles}}, \bibinfo {author}
  {\bibfnamefont {J.}~\bibnamefont {Marshall}}, \bibinfo {author}
  {\bibfnamefont {D.}~\bibnamefont {Scarabelli}}, \bibinfo {author}
  {\bibfnamefont {G.}~\bibnamefont {Crossman}}, \bibinfo {author}
  {\bibfnamefont {K.}~\bibnamefont {Yadav}}, \bibinfo {author} {\bibfnamefont
  {Y.}~\bibnamefont {Mohan}}, \bibinfo {author} {\bibfnamefont
  {C.}~\bibnamefont {Bui}}, \emph {et~al.},\ }\bibfield  {title} {\bibinfo
  {title} {Superconducting through-silicon vias for quantum integrated
  circuits},\ }\href@noop {} {\bibfield  {journal} {\bibinfo  {journal} {arXiv
  preprint arXiv:1708.02226}\ } (\bibinfo {year}
  {2017}{\natexlab{a}})}\BibitemShut {NoStop}%
\bibitem [{\citenamefont {Yost}\ \emph {et~al.}(2020)\citenamefont {Yost},
  \citenamefont {Schwartz}, \citenamefont {Mallek}, \citenamefont {Rosenberg},
  \citenamefont {Stull}, \citenamefont {Yoder}, \citenamefont {Calusine},
  \citenamefont {Cook}, \citenamefont {Das}, \citenamefont {Day} \emph
  {et~al.}}]{yost2020solid}%
  \BibitemOpen
  \bibfield  {author} {\bibinfo {author} {\bibfnamefont {D.-R.~W.}\
  \bibnamefont {Yost}}, \bibinfo {author} {\bibfnamefont {M.~E.}\ \bibnamefont
  {Schwartz}}, \bibinfo {author} {\bibfnamefont {J.}~\bibnamefont {Mallek}},
  \bibinfo {author} {\bibfnamefont {D.}~\bibnamefont {Rosenberg}}, \bibinfo
  {author} {\bibfnamefont {C.}~\bibnamefont {Stull}}, \bibinfo {author}
  {\bibfnamefont {J.~L.}\ \bibnamefont {Yoder}}, \bibinfo {author}
  {\bibfnamefont {G.}~\bibnamefont {Calusine}}, \bibinfo {author}
  {\bibfnamefont {M.}~\bibnamefont {Cook}}, \bibinfo {author} {\bibfnamefont
  {R.}~\bibnamefont {Das}}, \bibinfo {author} {\bibfnamefont {A.~L.}\
  \bibnamefont {Day}}, \emph {et~al.},\ }\bibfield  {title} {\bibinfo {title}
  {Solid-state qubits integrated with superconducting through-silicon vias},\
  }\href@noop {} {\bibfield  {journal} {\bibinfo  {journal} {npj Quantum
  Information}\ }\textbf {\bibinfo {volume} {6}},\ \bibinfo {pages} {1}
  (\bibinfo {year} {2020})}\BibitemShut {NoStop}%
\bibitem [{\citenamefont {Rahamim}\ \emph {et~al.}(2017)\citenamefont
  {Rahamim}, \citenamefont {Behrle}, \citenamefont {Peterer}, \citenamefont
  {Patterson}, \citenamefont {Spring}, \citenamefont {Tsunoda}, \citenamefont
  {Manenti}, \citenamefont {Tancredi},\ and\ \citenamefont
  {Leek}}]{rahamim2017double}%
  \BibitemOpen
  \bibfield  {author} {\bibinfo {author} {\bibfnamefont {J.}~\bibnamefont
  {Rahamim}}, \bibinfo {author} {\bibfnamefont {T.}~\bibnamefont {Behrle}},
  \bibinfo {author} {\bibfnamefont {M.}~\bibnamefont {Peterer}}, \bibinfo
  {author} {\bibfnamefont {A.}~\bibnamefont {Patterson}}, \bibinfo {author}
  {\bibfnamefont {P.}~\bibnamefont {Spring}}, \bibinfo {author} {\bibfnamefont
  {T.}~\bibnamefont {Tsunoda}}, \bibinfo {author} {\bibfnamefont
  {R.}~\bibnamefont {Manenti}}, \bibinfo {author} {\bibfnamefont
  {G.}~\bibnamefont {Tancredi}},\ and\ \bibinfo {author} {\bibfnamefont
  {P.}~\bibnamefont {Leek}},\ }\bibfield  {title} {\bibinfo {title}
  {Double-sided coaxial circuit qed with out-of-plane wiring},\ }\href@noop {}
  {\bibfield  {journal} {\bibinfo  {journal} {Applied Physics Letters}\
  }\textbf {\bibinfo {volume} {110}},\ \bibinfo {pages} {222602} (\bibinfo
  {year} {2017})}\BibitemShut {NoStop}%
\bibitem [{\citenamefont {Otterbach}\ \emph {et~al.}(2017)\citenamefont
  {Otterbach}, \citenamefont {Manenti}, \citenamefont {Alidoust}, \citenamefont
  {Bestwick}, \citenamefont {Block}, \citenamefont {Bloom}, \citenamefont
  {Caldwell}, \citenamefont {Didier}, \citenamefont {Fried}, \citenamefont
  {Hong} \emph {et~al.}}]{otterbach2017unsupervised}%
  \BibitemOpen
  \bibfield  {author} {\bibinfo {author} {\bibfnamefont {J.}~\bibnamefont
  {Otterbach}}, \bibinfo {author} {\bibfnamefont {R.}~\bibnamefont {Manenti}},
  \bibinfo {author} {\bibfnamefont {N.}~\bibnamefont {Alidoust}}, \bibinfo
  {author} {\bibfnamefont {A.}~\bibnamefont {Bestwick}}, \bibinfo {author}
  {\bibfnamefont {M.}~\bibnamefont {Block}}, \bibinfo {author} {\bibfnamefont
  {B.}~\bibnamefont {Bloom}}, \bibinfo {author} {\bibfnamefont
  {S.}~\bibnamefont {Caldwell}}, \bibinfo {author} {\bibfnamefont
  {N.}~\bibnamefont {Didier}}, \bibinfo {author} {\bibfnamefont {E.~S.}\
  \bibnamefont {Fried}}, \bibinfo {author} {\bibfnamefont {S.}~\bibnamefont
  {Hong}}, \emph {et~al.},\ }\bibfield  {title} {\bibinfo {title} {Unsupervised
  machine learning on a hybrid quantum computer},\ }\href@noop {} {\bibfield
  {journal} {\bibinfo  {journal} {arXiv preprint arXiv:1712.05771}\ } (\bibinfo
  {year} {2017})}\BibitemShut {NoStop}%
\bibitem [{\citenamefont {Jurcevic}\ \emph {et~al.}(2020)\citenamefont
  {Jurcevic}, \citenamefont {Javadi-Abhari}, \citenamefont {Bishop},
  \citenamefont {Lauer}, \citenamefont {Bogorin}, \citenamefont {Brink},
  \citenamefont {Capelluto}, \citenamefont {G{\"u}nl{\"u}k}, \citenamefont
  {Itoko}, \citenamefont {Kanazawa} \emph
  {et~al.}}]{jurcevic2020demonstration}%
  \BibitemOpen
  \bibfield  {author} {\bibinfo {author} {\bibfnamefont {P.}~\bibnamefont
  {Jurcevic}}, \bibinfo {author} {\bibfnamefont {A.}~\bibnamefont
  {Javadi-Abhari}}, \bibinfo {author} {\bibfnamefont {L.~S.}\ \bibnamefont
  {Bishop}}, \bibinfo {author} {\bibfnamefont {I.}~\bibnamefont {Lauer}},
  \bibinfo {author} {\bibfnamefont {D.~F.}\ \bibnamefont {Bogorin}}, \bibinfo
  {author} {\bibfnamefont {M.}~\bibnamefont {Brink}}, \bibinfo {author}
  {\bibfnamefont {L.}~\bibnamefont {Capelluto}}, \bibinfo {author}
  {\bibfnamefont {O.}~\bibnamefont {G{\"u}nl{\"u}k}}, \bibinfo {author}
  {\bibfnamefont {T.}~\bibnamefont {Itoko}}, \bibinfo {author} {\bibfnamefont
  {N.}~\bibnamefont {Kanazawa}}, \emph {et~al.},\ }\bibfield  {title} {\bibinfo
  {title} {Demonstration of quantum volume 64 on a superconducting quantum
  computing system},\ }\href@noop {} {\bibfield  {journal} {\bibinfo  {journal}
  {arXiv preprint arXiv:2008.08571}\ } (\bibinfo {year} {2020})}\BibitemShut
  {NoStop}%
\bibitem [{\citenamefont {Jerger}\ \emph {et~al.}(2012)\citenamefont {Jerger},
  \citenamefont {Poletto}, \citenamefont {Macha}, \citenamefont {H{\"u}bner},
  \citenamefont {Il’ichev},\ and\ \citenamefont
  {Ustinov}}]{jerger2012frequency}%
  \BibitemOpen
  \bibfield  {author} {\bibinfo {author} {\bibfnamefont {M.}~\bibnamefont
  {Jerger}}, \bibinfo {author} {\bibfnamefont {S.}~\bibnamefont {Poletto}},
  \bibinfo {author} {\bibfnamefont {P.}~\bibnamefont {Macha}}, \bibinfo
  {author} {\bibfnamefont {U.}~\bibnamefont {H{\"u}bner}}, \bibinfo {author}
  {\bibfnamefont {E.}~\bibnamefont {Il’ichev}},\ and\ \bibinfo {author}
  {\bibfnamefont {A.~V.}\ \bibnamefont {Ustinov}},\ }\bibfield  {title}
  {\bibinfo {title} {Frequency division multiplexing readout and simultaneous
  manipulation of an array of flux qubits},\ }\href@noop {} {\bibfield
  {journal} {\bibinfo  {journal} {Applied Physics Letters}\ }\textbf {\bibinfo
  {volume} {101}},\ \bibinfo {pages} {042604} (\bibinfo {year}
  {2012})}\BibitemShut {NoStop}%
\bibitem [{\citenamefont {Kelly}\ \emph {et~al.}(2015)\citenamefont {Kelly},
  \citenamefont {Barends}, \citenamefont {Fowler}, \citenamefont {Megrant},
  \citenamefont {Jeffrey}, \citenamefont {White}, \citenamefont {Sank},
  \citenamefont {Mutus}, \citenamefont {Campbell}, \citenamefont {Chen} \emph
  {et~al.}}]{kelly2015state}%
  \BibitemOpen
  \bibfield  {author} {\bibinfo {author} {\bibfnamefont {J.}~\bibnamefont
  {Kelly}}, \bibinfo {author} {\bibfnamefont {R.}~\bibnamefont {Barends}},
  \bibinfo {author} {\bibfnamefont {A.~G.}\ \bibnamefont {Fowler}}, \bibinfo
  {author} {\bibfnamefont {A.}~\bibnamefont {Megrant}}, \bibinfo {author}
  {\bibfnamefont {E.}~\bibnamefont {Jeffrey}}, \bibinfo {author} {\bibfnamefont
  {T.~C.}\ \bibnamefont {White}}, \bibinfo {author} {\bibfnamefont
  {D.}~\bibnamefont {Sank}}, \bibinfo {author} {\bibfnamefont {J.~Y.}\
  \bibnamefont {Mutus}}, \bibinfo {author} {\bibfnamefont {B.}~\bibnamefont
  {Campbell}}, \bibinfo {author} {\bibfnamefont {Y.}~\bibnamefont {Chen}},
  \emph {et~al.},\ }\bibfield  {title} {\bibinfo {title} {State preservation by
  repetitive error detection in a superconducting quantum circuit},\
  }\href@noop {} {\bibfield  {journal} {\bibinfo  {journal} {Nature}\ }\textbf
  {\bibinfo {volume} {519}},\ \bibinfo {pages} {66} (\bibinfo {year}
  {2015})}\BibitemShut {NoStop}%
\bibitem [{\citenamefont {Bardin}\ \emph {et~al.}(2021)\citenamefont {Bardin},
  \citenamefont {Slichter},\ and\ \citenamefont
  {Reilly}}]{bardin2021microwaves}%
  \BibitemOpen
  \bibfield  {author} {\bibinfo {author} {\bibfnamefont {J.~C.}\ \bibnamefont
  {Bardin}}, \bibinfo {author} {\bibfnamefont {D.~H.}\ \bibnamefont
  {Slichter}},\ and\ \bibinfo {author} {\bibfnamefont {D.~J.}\ \bibnamefont
  {Reilly}},\ }\bibfield  {title} {\bibinfo {title} {Microwaves in quantum
  computing},\ }\href@noop {} {\bibfield  {journal} {\bibinfo  {journal} {IEEE
  Journal of Microwaves}\ }\textbf {\bibinfo {volume} {1}},\ \bibinfo {pages}
  {403} (\bibinfo {year} {2021})}\BibitemShut {NoStop}%
\bibitem [{\citenamefont {Krupka}\ \emph {et~al.}(2006)\citenamefont {Krupka},
  \citenamefont {Breeze}, \citenamefont {Centeno}, \citenamefont {Alford},
  \citenamefont {Claussen},\ and\ \citenamefont
  {Jensen}}]{krupka2006measurements}%
  \BibitemOpen
  \bibfield  {author} {\bibinfo {author} {\bibfnamefont {J.}~\bibnamefont
  {Krupka}}, \bibinfo {author} {\bibfnamefont {J.}~\bibnamefont {Breeze}},
  \bibinfo {author} {\bibfnamefont {A.}~\bibnamefont {Centeno}}, \bibinfo
  {author} {\bibfnamefont {N.}~\bibnamefont {Alford}}, \bibinfo {author}
  {\bibfnamefont {T.}~\bibnamefont {Claussen}},\ and\ \bibinfo {author}
  {\bibfnamefont {L.}~\bibnamefont {Jensen}},\ }\bibfield  {title} {\bibinfo
  {title} {Measurements of permittivity, dielectric loss tangent, and
  resistivity of float-zone silicon at microwave frequencies},\ }\href@noop {}
  {\bibfield  {journal} {\bibinfo  {journal} {IEEE Transactions on microwave
  theory and techniques}\ }\textbf {\bibinfo {volume} {54}},\ \bibinfo {pages}
  {3995} (\bibinfo {year} {2006})}\BibitemShut {NoStop}%
\bibitem [{\citenamefont {Vahidpour}\ \emph
  {et~al.}(2017{\natexlab{b}})\citenamefont {Vahidpour}, \citenamefont
  {O'Brien}, \citenamefont {Whyland}, \citenamefont {Angeles}, \citenamefont
  {Marshall}, \citenamefont {Scarabelli}, \citenamefont {Crossman},
  \citenamefont {Yadav}, \citenamefont {Mohan}, \citenamefont {Bui},
  \citenamefont {Rawat}, \citenamefont {Renzas}, \citenamefont {Vodrahalli},
  \citenamefont {Bestwick},\ and\ \citenamefont {Rigetti}}]{Vahidpour2017}%
  \BibitemOpen
  \bibfield  {author} {\bibinfo {author} {\bibfnamefont {M.}~\bibnamefont
  {Vahidpour}}, \bibinfo {author} {\bibfnamefont {W.}~\bibnamefont {O'Brien}},
  \bibinfo {author} {\bibfnamefont {J.~T.}\ \bibnamefont {Whyland}}, \bibinfo
  {author} {\bibfnamefont {J.}~\bibnamefont {Angeles}}, \bibinfo {author}
  {\bibfnamefont {J.}~\bibnamefont {Marshall}}, \bibinfo {author}
  {\bibfnamefont {D.}~\bibnamefont {Scarabelli}}, \bibinfo {author}
  {\bibfnamefont {G.}~\bibnamefont {Crossman}}, \bibinfo {author}
  {\bibfnamefont {K.}~\bibnamefont {Yadav}}, \bibinfo {author} {\bibfnamefont
  {Y.}~\bibnamefont {Mohan}}, \bibinfo {author} {\bibfnamefont
  {C.}~\bibnamefont {Bui}}, \bibinfo {author} {\bibfnamefont {V.}~\bibnamefont
  {Rawat}}, \bibinfo {author} {\bibfnamefont {R.}~\bibnamefont {Renzas}},
  \bibinfo {author} {\bibfnamefont {N.}~\bibnamefont {Vodrahalli}}, \bibinfo
  {author} {\bibfnamefont {A.}~\bibnamefont {Bestwick}},\ and\ \bibinfo
  {author} {\bibfnamefont {C.}~\bibnamefont {Rigetti}},\ }\bibfield  {title}
  {\bibinfo {title} {Superconducting through-silicon vias for quantum
  integrated circuits},\ }\href {https://arxiv.org/abs/1708.02226} {\bibfield
  {journal} {\bibinfo  {journal} {arXiv:1708.02226}\ } (\bibinfo {year}
  {2017}{\natexlab{b}})}\BibitemShut {NoStop}%
\bibitem [{\citenamefont {O'Brien}\ \emph {et~al.}(2017)\citenamefont
  {O'Brien}, \citenamefont {Vahidpour}, \citenamefont {Whyland}, \citenamefont
  {Angeles}, \citenamefont {Marshall}, \citenamefont {Scarabelli},
  \citenamefont {Crossman}, \citenamefont {Yadav}, \citenamefont {Mohan},
  \citenamefont {Bui} \emph {et~al.}}]{o2017superconducting}%
  \BibitemOpen
  \bibfield  {author} {\bibinfo {author} {\bibfnamefont {W.}~\bibnamefont
  {O'Brien}}, \bibinfo {author} {\bibfnamefont {M.}~\bibnamefont {Vahidpour}},
  \bibinfo {author} {\bibfnamefont {J.~T.}\ \bibnamefont {Whyland}}, \bibinfo
  {author} {\bibfnamefont {J.}~\bibnamefont {Angeles}}, \bibinfo {author}
  {\bibfnamefont {J.}~\bibnamefont {Marshall}}, \bibinfo {author}
  {\bibfnamefont {D.}~\bibnamefont {Scarabelli}}, \bibinfo {author}
  {\bibfnamefont {G.}~\bibnamefont {Crossman}}, \bibinfo {author}
  {\bibfnamefont {K.}~\bibnamefont {Yadav}}, \bibinfo {author} {\bibfnamefont
  {Y.}~\bibnamefont {Mohan}}, \bibinfo {author} {\bibfnamefont
  {C.}~\bibnamefont {Bui}}, \emph {et~al.},\ }\bibfield  {title} {\bibinfo
  {title} {Superconducting caps for quantum integrated circuits},\ }\href@noop
  {} {\bibfield  {journal} {\bibinfo  {journal} {arXiv preprint
  arXiv:1708.02219}\ } (\bibinfo {year} {2017})}\BibitemShut {NoStop}%
\bibitem [{\citenamefont {Krinner}\ \emph {et~al.}(2019)\citenamefont
  {Krinner}, \citenamefont {Storz}, \citenamefont {Kurpiers}, \citenamefont
  {Magnard}, \citenamefont {Heinsoo}, \citenamefont {Keller}, \citenamefont
  {Luetolf}, \citenamefont {Eichler},\ and\ \citenamefont
  {Wallraff}}]{krinner2019engineering}%
  \BibitemOpen
  \bibfield  {author} {\bibinfo {author} {\bibfnamefont {S.}~\bibnamefont
  {Krinner}}, \bibinfo {author} {\bibfnamefont {S.}~\bibnamefont {Storz}},
  \bibinfo {author} {\bibfnamefont {P.}~\bibnamefont {Kurpiers}}, \bibinfo
  {author} {\bibfnamefont {P.}~\bibnamefont {Magnard}}, \bibinfo {author}
  {\bibfnamefont {J.}~\bibnamefont {Heinsoo}}, \bibinfo {author} {\bibfnamefont
  {R.}~\bibnamefont {Keller}}, \bibinfo {author} {\bibfnamefont
  {J.}~\bibnamefont {Luetolf}}, \bibinfo {author} {\bibfnamefont
  {C.}~\bibnamefont {Eichler}},\ and\ \bibinfo {author} {\bibfnamefont
  {A.}~\bibnamefont {Wallraff}},\ }\bibfield  {title} {\bibinfo {title}
  {Engineering cryogenic setups for 100-qubit scale superconducting circuit
  systems},\ }\href@noop {} {\bibfield  {journal} {\bibinfo  {journal} {EPJ
  Quantum Technology}\ }\textbf {\bibinfo {volume} {6}},\ \bibinfo {pages} {2}
  (\bibinfo {year} {2019})}\BibitemShut {NoStop}%
\bibitem [{\citenamefont {Caldwell}\ \emph {et~al.}(2018)\citenamefont
  {Caldwell}, \citenamefont {Didier}, \citenamefont {Ryan}, \citenamefont
  {Sete}, \citenamefont {Hudson}, \citenamefont {Karalekas}, \citenamefont
  {Manenti}, \citenamefont {da~Silva}, \citenamefont {Sinclair}, \citenamefont
  {Acala} \emph {et~al.}}]{caldwell2018parametrically}%
  \BibitemOpen
  \bibfield  {author} {\bibinfo {author} {\bibfnamefont {S.}~\bibnamefont
  {Caldwell}}, \bibinfo {author} {\bibfnamefont {N.}~\bibnamefont {Didier}},
  \bibinfo {author} {\bibfnamefont {C.}~\bibnamefont {Ryan}}, \bibinfo {author}
  {\bibfnamefont {E.}~\bibnamefont {Sete}}, \bibinfo {author} {\bibfnamefont
  {A.}~\bibnamefont {Hudson}}, \bibinfo {author} {\bibfnamefont
  {P.}~\bibnamefont {Karalekas}}, \bibinfo {author} {\bibfnamefont
  {R.}~\bibnamefont {Manenti}}, \bibinfo {author} {\bibfnamefont
  {M.}~\bibnamefont {da~Silva}}, \bibinfo {author} {\bibfnamefont
  {R.}~\bibnamefont {Sinclair}}, \bibinfo {author} {\bibfnamefont
  {E.}~\bibnamefont {Acala}}, \emph {et~al.},\ }\bibfield  {title} {\bibinfo
  {title} {Parametrically activated entangling gates using transmon qubits},\
  }\href@noop {} {\bibfield  {journal} {\bibinfo  {journal} {Physical Review
  Applied}\ }\textbf {\bibinfo {volume} {10}},\ \bibinfo {pages} {034050}
  (\bibinfo {year} {2018})}\BibitemShut {NoStop}%
\bibitem [{\citenamefont {C{\'o}rcoles}\ \emph {et~al.}(2011)\citenamefont
  {C{\'o}rcoles}, \citenamefont {Chow}, \citenamefont {Gambetta}, \citenamefont
  {Rigetti}, \citenamefont {Rozen}, \citenamefont {Keefe}, \citenamefont
  {Beth~Rothwell}, \citenamefont {Ketchen},\ and\ \citenamefont
  {Steffen}}]{corcoles2011protecting}%
  \BibitemOpen
  \bibfield  {author} {\bibinfo {author} {\bibfnamefont {A.~D.}\ \bibnamefont
  {C{\'o}rcoles}}, \bibinfo {author} {\bibfnamefont {J.~M.}\ \bibnamefont
  {Chow}}, \bibinfo {author} {\bibfnamefont {J.~M.}\ \bibnamefont {Gambetta}},
  \bibinfo {author} {\bibfnamefont {C.}~\bibnamefont {Rigetti}}, \bibinfo
  {author} {\bibfnamefont {J.~R.}\ \bibnamefont {Rozen}}, \bibinfo {author}
  {\bibfnamefont {G.~A.}\ \bibnamefont {Keefe}}, \bibinfo {author}
  {\bibfnamefont {M.}~\bibnamefont {Beth~Rothwell}}, \bibinfo {author}
  {\bibfnamefont {M.~B.}\ \bibnamefont {Ketchen}},\ and\ \bibinfo {author}
  {\bibfnamefont {M.}~\bibnamefont {Steffen}},\ }\bibfield  {title} {\bibinfo
  {title} {Protecting superconducting qubits from radiation},\ }\href@noop {}
  {\bibfield  {journal} {\bibinfo  {journal} {Applied Physics Letters}\
  }\textbf {\bibinfo {volume} {99}},\ \bibinfo {pages} {181906} (\bibinfo
  {year} {2011})}\BibitemShut {NoStop}%
\bibitem [{\citenamefont {Barends}\ \emph {et~al.}(2011)\citenamefont
  {Barends}, \citenamefont {Wenner}, \citenamefont {Lenander}, \citenamefont
  {Chen}, \citenamefont {Bialczak}, \citenamefont {Kelly}, \citenamefont
  {Lucero}, \citenamefont {O’Malley}, \citenamefont {Mariantoni},
  \citenamefont {Sank} \emph {et~al.}}]{barends2011minimizing}%
  \BibitemOpen
  \bibfield  {author} {\bibinfo {author} {\bibfnamefont {R.}~\bibnamefont
  {Barends}}, \bibinfo {author} {\bibfnamefont {J.}~\bibnamefont {Wenner}},
  \bibinfo {author} {\bibfnamefont {M.}~\bibnamefont {Lenander}}, \bibinfo
  {author} {\bibfnamefont {Y.}~\bibnamefont {Chen}}, \bibinfo {author}
  {\bibfnamefont {R.~C.}\ \bibnamefont {Bialczak}}, \bibinfo {author}
  {\bibfnamefont {J.}~\bibnamefont {Kelly}}, \bibinfo {author} {\bibfnamefont
  {E.}~\bibnamefont {Lucero}}, \bibinfo {author} {\bibfnamefont
  {P.}~\bibnamefont {O’Malley}}, \bibinfo {author} {\bibfnamefont
  {M.}~\bibnamefont {Mariantoni}}, \bibinfo {author} {\bibfnamefont
  {D.}~\bibnamefont {Sank}}, \emph {et~al.},\ }\bibfield  {title} {\bibinfo
  {title} {Minimizing quasiparticle generation from stray infrared light in
  superconducting quantum circuits},\ }\href@noop {} {\bibfield  {journal}
  {\bibinfo  {journal} {Applied Physics Letters}\ }\textbf {\bibinfo {volume}
  {99}},\ \bibinfo {pages} {113507} (\bibinfo {year} {2011})}\BibitemShut
  {NoStop}%
\bibitem [{\citenamefont {Braum{\"u}ller}\ \emph {et~al.}(2020)\citenamefont
  {Braum{\"u}ller}, \citenamefont {Ding}, \citenamefont {Veps{\"a}l{\"a}inen},
  \citenamefont {Sung}, \citenamefont {Kjaergaard}, \citenamefont {Menke},
  \citenamefont {Winik}, \citenamefont {Kim}, \citenamefont {Niedzielski},
  \citenamefont {Melville} \emph {et~al.}}]{braumuller2020characterizing}%
  \BibitemOpen
  \bibfield  {author} {\bibinfo {author} {\bibfnamefont {J.}~\bibnamefont
  {Braum{\"u}ller}}, \bibinfo {author} {\bibfnamefont {L.}~\bibnamefont
  {Ding}}, \bibinfo {author} {\bibfnamefont {A.~P.}\ \bibnamefont
  {Veps{\"a}l{\"a}inen}}, \bibinfo {author} {\bibfnamefont {Y.}~\bibnamefont
  {Sung}}, \bibinfo {author} {\bibfnamefont {M.}~\bibnamefont {Kjaergaard}},
  \bibinfo {author} {\bibfnamefont {T.}~\bibnamefont {Menke}}, \bibinfo
  {author} {\bibfnamefont {R.}~\bibnamefont {Winik}}, \bibinfo {author}
  {\bibfnamefont {D.}~\bibnamefont {Kim}}, \bibinfo {author} {\bibfnamefont
  {B.~M.}\ \bibnamefont {Niedzielski}}, \bibinfo {author} {\bibfnamefont
  {A.}~\bibnamefont {Melville}}, \emph {et~al.},\ }\bibfield  {title} {\bibinfo
  {title} {Characterizing and optimizing qubit coherence based on squid
  geometry},\ }\href@noop {} {\bibfield  {journal} {\bibinfo  {journal}
  {Physical Review Applied}\ }\textbf {\bibinfo {volume} {13}},\ \bibinfo
  {pages} {054079} (\bibinfo {year} {2020})}\BibitemShut {NoStop}%
\bibitem [{\citenamefont {Burnett}\ \emph {et~al.}(2019)\citenamefont
  {Burnett}, \citenamefont {Bengtsson}, \citenamefont {Scigliuzzo},
  \citenamefont {Niepce}, \citenamefont {Kudra}, \citenamefont {Delsing},\ and\
  \citenamefont {Bylander}}]{burnett2019decoherence}%
  \BibitemOpen
  \bibfield  {author} {\bibinfo {author} {\bibfnamefont {J.~J.}\ \bibnamefont
  {Burnett}}, \bibinfo {author} {\bibfnamefont {A.}~\bibnamefont {Bengtsson}},
  \bibinfo {author} {\bibfnamefont {M.}~\bibnamefont {Scigliuzzo}}, \bibinfo
  {author} {\bibfnamefont {D.}~\bibnamefont {Niepce}}, \bibinfo {author}
  {\bibfnamefont {M.}~\bibnamefont {Kudra}}, \bibinfo {author} {\bibfnamefont
  {P.}~\bibnamefont {Delsing}},\ and\ \bibinfo {author} {\bibfnamefont
  {J.}~\bibnamefont {Bylander}},\ }\bibfield  {title} {\bibinfo {title}
  {Decoherence benchmarking of superconducting qubits},\ }\href@noop {}
  {\bibfield  {journal} {\bibinfo  {journal} {npj Quantum Information}\
  }\textbf {\bibinfo {volume} {5}},\ \bibinfo {pages} {1} (\bibinfo {year}
  {2019})}\BibitemShut {NoStop}%
\bibitem [{\citenamefont {Schl{\"o}r}\ \emph {et~al.}(2019)\citenamefont
  {Schl{\"o}r}, \citenamefont {Lisenfeld}, \citenamefont {M{\"u}ller},
  \citenamefont {Bilmes}, \citenamefont {Schneider}, \citenamefont {Pappas},
  \citenamefont {Ustinov},\ and\ \citenamefont
  {Weides}}]{schlor2019correlating}%
  \BibitemOpen
  \bibfield  {author} {\bibinfo {author} {\bibfnamefont {S.}~\bibnamefont
  {Schl{\"o}r}}, \bibinfo {author} {\bibfnamefont {J.}~\bibnamefont
  {Lisenfeld}}, \bibinfo {author} {\bibfnamefont {C.}~\bibnamefont
  {M{\"u}ller}}, \bibinfo {author} {\bibfnamefont {A.}~\bibnamefont {Bilmes}},
  \bibinfo {author} {\bibfnamefont {A.}~\bibnamefont {Schneider}}, \bibinfo
  {author} {\bibfnamefont {D.~P.}\ \bibnamefont {Pappas}}, \bibinfo {author}
  {\bibfnamefont {A.~V.}\ \bibnamefont {Ustinov}},\ and\ \bibinfo {author}
  {\bibfnamefont {M.}~\bibnamefont {Weides}},\ }\bibfield  {title} {\bibinfo
  {title} {Correlating decoherence in transmon qubits: Low frequency noise by
  single fluctuators},\ }\href@noop {} {\bibfield  {journal} {\bibinfo
  {journal} {Physical Rev. Lett.}\ }\textbf {\bibinfo {volume} {123}},\
  \bibinfo {pages} {190502} (\bibinfo {year} {2019})}\BibitemShut {NoStop}%
\bibitem [{\citenamefont {Asaad}\ \emph {et~al.}(2016)\citenamefont {Asaad},
  \citenamefont {Dickel}, \citenamefont {Langford}, \citenamefont {Poletto},
  \citenamefont {Bruno}, \citenamefont {Rol}, \citenamefont {Deurloo},\ and\
  \citenamefont {DiCarlo}}]{asaad2016independent}%
  \BibitemOpen
  \bibfield  {author} {\bibinfo {author} {\bibfnamefont {S.}~\bibnamefont
  {Asaad}}, \bibinfo {author} {\bibfnamefont {C.}~\bibnamefont {Dickel}},
  \bibinfo {author} {\bibfnamefont {N.~K.}\ \bibnamefont {Langford}}, \bibinfo
  {author} {\bibfnamefont {S.}~\bibnamefont {Poletto}}, \bibinfo {author}
  {\bibfnamefont {A.}~\bibnamefont {Bruno}}, \bibinfo {author} {\bibfnamefont
  {M.~A.}\ \bibnamefont {Rol}}, \bibinfo {author} {\bibfnamefont
  {D.}~\bibnamefont {Deurloo}},\ and\ \bibinfo {author} {\bibfnamefont
  {L.}~\bibnamefont {DiCarlo}},\ }\bibfield  {title} {\bibinfo {title}
  {Independent, extensible control of same-frequency superconducting qubits by
  selective broadcasting},\ }\href@noop {} {\bibfield  {journal} {\bibinfo
  {journal} {npj Quantum Information}\ }\textbf {\bibinfo {volume} {2}},\
  \bibinfo {pages} {1} (\bibinfo {year} {2016})}\BibitemShut {NoStop}%
\bibitem [{\citenamefont {McKay}\ \emph {et~al.}(2016)\citenamefont {McKay},
  \citenamefont {Filipp}, \citenamefont {Mezzacapo}, \citenamefont {Magesan},
  \citenamefont {Chow},\ and\ \citenamefont {Gambetta}}]{mckay2016universal}%
  \BibitemOpen
  \bibfield  {author} {\bibinfo {author} {\bibfnamefont {D.~C.}\ \bibnamefont
  {McKay}}, \bibinfo {author} {\bibfnamefont {S.}~\bibnamefont {Filipp}},
  \bibinfo {author} {\bibfnamefont {A.}~\bibnamefont {Mezzacapo}}, \bibinfo
  {author} {\bibfnamefont {E.}~\bibnamefont {Magesan}}, \bibinfo {author}
  {\bibfnamefont {J.~M.}\ \bibnamefont {Chow}},\ and\ \bibinfo {author}
  {\bibfnamefont {J.~M.}\ \bibnamefont {Gambetta}},\ }\bibfield  {title}
  {\bibinfo {title} {Universal gate for fixed-frequency qubits via a tunable
  bus},\ }\href@noop {} {\bibfield  {journal} {\bibinfo  {journal} {Physical
  Review Applied}\ }\textbf {\bibinfo {volume} {6}},\ \bibinfo {pages} {064007}
  (\bibinfo {year} {2016})}\BibitemShut {NoStop}%
\bibitem [{\citenamefont {Didier}\ \emph {et~al.}(2018)\citenamefont {Didier},
  \citenamefont {Sete}, \citenamefont {da~Silva},\ and\ \citenamefont
  {Rigetti}}]{Didier2017}%
  \BibitemOpen
  \bibfield  {author} {\bibinfo {author} {\bibfnamefont {N.}~\bibnamefont
  {Didier}}, \bibinfo {author} {\bibfnamefont {E.~A.}\ \bibnamefont {Sete}},
  \bibinfo {author} {\bibfnamefont {M.~P.}\ \bibnamefont {da~Silva}},\ and\
  \bibinfo {author} {\bibfnamefont {C.}~\bibnamefont {Rigetti}},\ }\bibfield
  {title} {\bibinfo {title} {Analytical modeling of parametrically modulated
  transmon qubits},\ }\href@noop {} {\bibfield  {journal} {\bibinfo  {journal}
  {Physical Review A}\ }\textbf {\bibinfo {volume} {97}},\ \bibinfo {pages}
  {022330} (\bibinfo {year} {2018})}\BibitemShut {NoStop}%
\bibitem [{\citenamefont {Didier}\ \emph {et~al.}(2019)\citenamefont {Didier},
  \citenamefont {Sete}, \citenamefont {Combes},\ and\ \citenamefont
  {da~Silva}}]{Didier2019}%
  \BibitemOpen
  \bibfield  {author} {\bibinfo {author} {\bibfnamefont {N.}~\bibnamefont
  {Didier}}, \bibinfo {author} {\bibfnamefont {E.~A.}\ \bibnamefont {Sete}},
  \bibinfo {author} {\bibfnamefont {J.}~\bibnamefont {Combes}},\ and\ \bibinfo
  {author} {\bibfnamefont {M.~P.}\ \bibnamefont {da~Silva}},\ }\bibfield
  {title} {\bibinfo {title} {ac flux sweet spots in parametrically modulated
  superconducting qubits},\ }\href
  {https://doi.org/10.1103/PhysRevApplied.12.054015} {\bibfield  {journal}
  {\bibinfo  {journal} {Phys. Rev. Applied}\ }\textbf {\bibinfo {volume}
  {12}},\ \bibinfo {pages} {054015} (\bibinfo {year} {2019})}\BibitemShut
  {NoStop}%
\bibitem [{\citenamefont {Lecocq}\ \emph {et~al.}(2021)\citenamefont {Lecocq},
  \citenamefont {Quinlan}, \citenamefont {Cicak}, \citenamefont {Aumentado},
  \citenamefont {Diddams},\ and\ \citenamefont {Teufel}}]{lecocq2021control}%
  \BibitemOpen
  \bibfield  {author} {\bibinfo {author} {\bibfnamefont {F.}~\bibnamefont
  {Lecocq}}, \bibinfo {author} {\bibfnamefont {F.}~\bibnamefont {Quinlan}},
  \bibinfo {author} {\bibfnamefont {K.}~\bibnamefont {Cicak}}, \bibinfo
  {author} {\bibfnamefont {J.}~\bibnamefont {Aumentado}}, \bibinfo {author}
  {\bibfnamefont {S.}~\bibnamefont {Diddams}},\ and\ \bibinfo {author}
  {\bibfnamefont {J.}~\bibnamefont {Teufel}},\ }\bibfield  {title} {\bibinfo
  {title} {Control and readout of a superconducting qubit using a photonic
  link},\ }\href@noop {} {\bibfield  {journal} {\bibinfo  {journal} {Nature}\
  }\textbf {\bibinfo {volume} {591}},\ \bibinfo {pages} {575} (\bibinfo {year}
  {2021})}\BibitemShut {NoStop}%
\end{thebibliography}%

\clearpage

%%%%%%%%%%
%%%%%%%%%%
%%%%%%%%%%
%%%%%%%%%%
%%%%%%%%%%

%%%%%%%%%% Merge with supplemental materials %%%%%%%%%%
\pagebreak

\widetext

\begin{center}

\section{\large Supplementary Information: \\XYZ lines for superconducting devices}
Riccardo Manenti,
Eyob A. Sete,
Angela Q. Chen,
Shobhan Kulshreshtha,
Jen-Hao Yeh,
Feyza Oruc,
Andrew Bestwick,
Mark Field,
Keith Jackson,
Stefano Poletto.\\
\emph{Rigetti Computing, Inc., Berkeley, CA}

\end{center}

%%%%%%%%%% Merge with supplemental materials %%%%%%%%%%
%%%%%%%%%% Prefix an "S" to all equations, figures, tables and reset the counter %%%%%%%%%%
\setcounter{equation}{0}
\setcounter{figure}{0}
\setcounter{table}{0}
\setcounter{page}{1}
\makeatletter
\renewcommand{\theequation}{S\arabic{equation}}
\renewcommand{\thefigure}{S\arabic{figure}}
\renewcommand{\thetable}{S\arabic{table}}
\renewcommand{\bibnumfmt}[1]{[#1]}
\renewcommand{\citenumfont}[1]{#1}
%%%%%%%%%% Prefix a "S" to all equations, figures, tables and reset the counter %%%%%%%%%%

\subsection{S1. Fridge setup and device parameters}

\begin{figure}[h!]
\includegraphics[width=0.8\columnwidth]{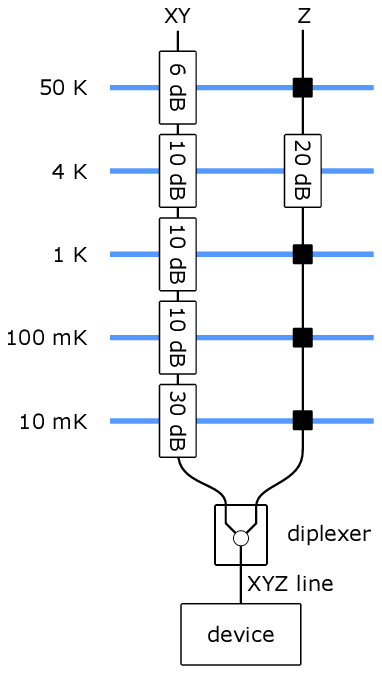}
\caption{
\textbf{Fridge setup.} 
Diagram of the fridge setup. The black squares represent SMA connectors. These components ensure a good thermalisation of the coaxial cable.
\label{fig:fridge}
}
\end{figure}

The cables connecting the room temperature instrumentation to the device must be suitably filtered and attenuated. Figure~\ref{fig:fridge}a  shows the fridge setup used in our work. 
The XY line has an overall attenuation of $66\;\text{dB}$ at DC. The Z line has only one $20\;\text{dB}$ attenuator installed at the 4~K plate. 
The diplexer shown at the bottom of the figure includes both microwave filters and eccosorb filters. 

\begin{figure}[h!]
\includegraphics[width=0.8\columnwidth]{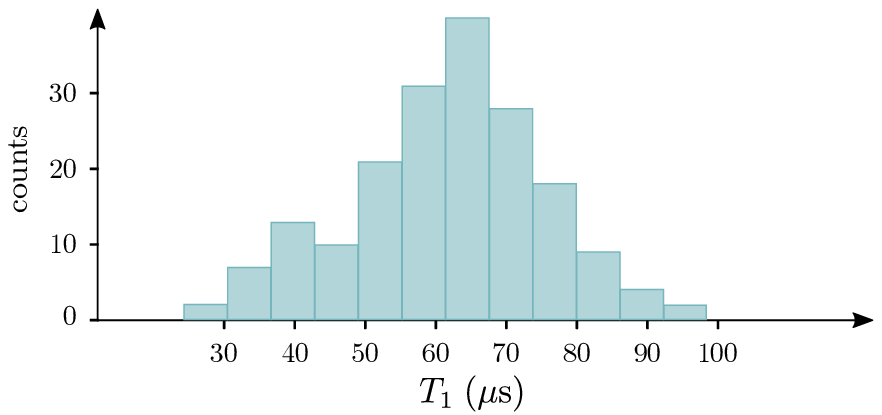}
\caption{
\textbf{Repeated measurements of $T_1$.} 
Histogram of 60 measurements of the relaxation time of qubit 2 in the course of a day.
\label{fig:T1-histogram}
}
\end{figure}

Table~\ref{table_suppl} reports the main device parameters. 
Here, $f_{\text{r}}$ is the resonator frequency,  $f^{\text{max}}_{\text{q}}$ ($f^{\text{min}}_{\text{q}}$) is the maximum (minimum) qubit frequency
and $\eta$ is the qubit anharmonicity. 
The coherence times $\tilde T_1$, $\tilde T^*_2$, and $\tilde T_{2E}$ are the median values measured at $f^{\text{max}}_{\text{q}}$. It should be pointed out that these values fluctuate in the course of a day (See Fig. \ref{fig:T1-histogram}. This phenomenon is well-known in the community and has been investigated in Ref.~\cite{burnett2019decoherence, schlor2019correlating}).
The figure of merit $F_{1\text{q}}$ is the fidelity of $100\;\text{ns}$ single-qubit gates measured with randomized benchmarking.

\begin{table}
\begin{center}%
\begin{tabular}{c|cccc}
 & q0 & q1 & q2 & q3\tabularnewline
\hline 
$f_{\text{r}}$ (MHz) & 7029 & 7122 & 7203 & 7292\tabularnewline
$f_{\text{q}}^{\text{max}}$ (MHz) & 3851 & 3786 & 3919 & 3870\tabularnewline
$f_{\text{q}}^{\text{min}}$ (MHz) & 2981 & 2956 & 3050 & 2936\tabularnewline
$\eta$ (MHz) & -206 & -208 & -208 & -210\tabularnewline
$E_{\text{C}}$ (MHz) & 182 & 185 & 185 & 186\tabularnewline
$E_{\text{J1}}$ (MHz) & 2140 & 2360 & 2160 & 2250\tabularnewline
$E_{\text{J2}}$ (MHz) & 9040 & 9090 & 9300 & 8860\tabularnewline
$\tilde{T}_{1}$ ($\mu\text{s}$) & 51 & 21 & 61 & 56\tabularnewline
$\tilde{T}_{2}^{*}$ ($\mu\text{s}$) & 10 & 16 & 10 & 1\tabularnewline
$\tilde{T}_{2\text{E}}$ ($\mu\text{s}$) & 81 & 25 & 60 & 31\tabularnewline
$F_{1\text{q}}$ (\%) & 99.93 & 99.77 & 99.88 & 98.09\tabularnewline
\end{tabular}\end{center}
\caption{Main parameters of the 4-qubit device investigated in this study.
\label{table_suppl}}
\end{table}

\pagebreak

\subsection{S2. Spurious qubit frequency modulation}

Since in our fridge configuration the XY fridge line and the Z fridge line
are combined together with a cryogenic diplexer,
 a control pulse at the qubit frequency sent through the XY fridge line can inadvertently modulate the qubit frequency.
Let us
assume that a $\pi$ pulse has a room
temperature voltage of $V(t)=A_{\text p}\cos(\omega_{\text{d}}t)$. This pulse
generates a current through the on-chip XYZ line of
\[
I(t)=2\alpha\frac{V(t)}{R},\qquad\qquad\qquad\alpha=10^{-\gamma/20\text{ dB}},
\]
where $\gamma$ is the attenuation of the line at the qubit frequency
(in our case, $\gamma\approx85\;\text{dB}$), and $R = 50 \;\Omega$ is the impedance
of the line. The factor of two takes into account the fact that the line is shorted to ground. The current produces a magnetic flux through the SQUID
of
\[
\Phi(t)=MI(t)=M2\alpha\frac{V(t)}{R},
\]
where $M$ is the mutual inductance between the line and the SQUID
(for our device, $M\approx500\;\text{fH}$). In our device, the amplitude required
to implement a $\pi$ pulse is $A_{\text p}=0.3\;V$ for a $100\;\text{ns}$
pulse. Thus, the magnetic flux threading the SQUID is
\[
\Phi(t)=M2\alpha\frac{A_{\text p}}{R}\cos(\omega_{\text{d}}t)=\Phi_{\text{ac}}\cos(\omega_{\text{d}}t),\qquad\text{where}\qquad\Phi_{\text{ac}}=1.6\cdot10^{-4}\Phi_{0}.
\]
Using Eq. (2) of the main text and the parameters for qubit 0, we conclude that the qubit frequency
would shift by $79\;\text{Hz}$. This shift is well below the qubit linewidth and cannot be detected with Ramsey experiments.

\subsection{S3. Ramsey experiments with flux pulses}

In this Section, we present the fitting function of the effective qubit frequency for the Ramsey experiment including flux pulses (Eq. (1) of the main text). 
Since the detailed derivation can be found in Ref.~\cite{Didier2017}, we limit ourselves to report the final result. First, let us define the vector
\[
\mathbf{c}=\Big \{4,-1,-\frac{1}{2^{2}},-\frac{21}{2^{7}},-\frac{19}{2^{7}},-\frac{5319}{2^{15}},-\frac{6649}{2^{15}},-\frac{1180581}{2^{22}},-\frac{446287}{2^{20}}\Big\},
\]
and the constants
\[
\xi=\sqrt{\frac{2\text{\ensuremath{E_{C}}}}{\sqrt{E_{\text{J}1}^{2}+E_{\text{J}2}^{2}}}},\qquad\qquad
\tilde{E}_{\text{J}}=\frac{2E_{\text{J}1}E_{\text{J}2}}{E_{\text{J}1}^{2}+E_{\text{J}2}^{2}}.
\]
We also need to introduce some coefficients $s_n$.
For $n=0$, we have
\[
s_{0}=E_{\text{C}}\sum_{j}c_{j}\xi^{j-2}\;_{2}F_{1}\Big[\frac{j-2}{8},\frac{j-2}{8}+\frac{1}{2},1,\tilde{E}_{j}^{2}\Big],
\]
where $_{2}F_{1}(a,b,c,d)$ is the hypergeometric function.
For $n>1$, we have
\[
s_{n}=\frac{2}{n!}E_{\text{C}}(-\tilde{E}_{\text{J}}/2)^{n}\sum_{j}c_{j}\xi^{j-2}\frac{\Gamma\big(n+\frac{j-2}{4}\big)}{\Gamma\big(\frac{j-2}{4}\big)}\;_{2}F_{1}\Big[\frac{n}{2}+\frac{j-2}{8},\frac{n+1}{2}+\frac{j-2}{8},n+1,\tilde{E}_{j}^{2}\Big],
\]
where $\Gamma(z)$ is the Euler gamma function. The effective qubit frequency is given by
\[
\bar{f}_{\text{q}}(\Phi_{\text{AC}})
=
\sum_{n=0}^{p}s_{n}\cos(2\pi n\Phi_{\text{dc}})J_{0}(2\pi n\Phi_{\text{ac}}),
\]
where $p$ defines the accuracy of the perturbation.
\end{document}